\documentclass[
reprint,
bibnotes,
amsmath,amssymb,
aps,
prd
]{revtex4-2}

\usepackage{graphicx}
\usepackage{dcolumn}
\usepackage{enumitem}
\usepackage{bm}
\usepackage[many]{tcolorbox}
\usepackage{shuffle}
\usepackage{physics}
\usepackage{tabularx}
\usepackage{diagbox}
\usepackage{subcaption}
\usepackage{multirow}
\usepackage{caption}
\usepackage{tikz}
\usetikzlibrary{decorations.markings}
\usetikzlibrary{decorations.pathmorphing}
\usetikzlibrary{intersections}
\usetikzlibrary{calc}
\usepackage{verbatim}
\RequirePackage[usenames,dvipsnames]{xcolor}

\usepackage{CJK}

\begin{document}

\begin{CJK*}{UTF8}{}
\CJKfamily{gbsn}

\title{Geometry of in-in correlators}

\author{Ross Glew}
\email{r.glew@herts.ac.uk}

\affiliation{
Department of Physics, Astronomy and Mathematics, University of Hertfordshire, Hatfield, Hertfordshire, AL10 9AB, UK\\
}

\begin{abstract}
We introduce a family of polytopes---{\it in-in zonotopes}---whose boundary structure organizes the contributions to scalar equal-time correlators in flat space computed via the in-in formalism. We provide explicit Minkowski sum and facet descriptions of these polytopes, and show that their boundaries factorize into products of graphical zonotopes and lower-dimensional in-in zonotopes, thereby mimicking the factorization structure of the correlators themselves. Evaluating their canonical forms at the origin---equivalently, calculating the volume of the dual polytope---reproduces the correlator. Finally, in a simple example, we show that the wavefunction decomposition of the correlator corresponds to a subdivision of the dual polytope.
\end{abstract}
\maketitle
\end{CJK*}

\section{Introduction}
Diagrammatic expansions play a central role in many quantum field theory computations. In recent years, significant progress has been made by taking the combinatorics of these diagrams seriously and asking whether they admit an underlying geometric interpretation. For instance, scattering amplitudes in $\text{tr}(\phi^3)$ theory are expressed as sums over Feynman diagrams; these diagrams label the vertices of the ABHY associahedron \cite{ABHY}. A more sophisticated version of this correspondence arises in $\mathcal{N}=4$ SYM, where on-shell diagrams assemble to tile the amplituhedron \cite{Arkani-Hamed:2013jha,Arkani-Hamed:2012zlh}. 

More recently, these ideas have been successfully extended into a cosmological setting: first through cosmological polytopes, which encode individual Feynman-diagram contributions to the so-called wavefunction of the universe \cite{CosmoPolys}, and more recently through the discovery of cosmohedra, which reorganize the entire sum over Feynman diagrams into a single polytope \cite{Cosmohedron}.

In this work, we consider another application of these ideas: equal-time correlators for a scalar theory with polynomial interactions in flat space. We focus on the combinatorial structure of diagrams appearing in the in-in expansion of the correlator.  To capture this structure, we introduce a polytope---the {\it in-in zonotope}---whose face poset organizes the diagrams graph by graph. We provide explicit Minkowski sum and facet descriptions of these polytopes, and show that their boundaries factorize into products of graphical zonotopes and lower-dimensional in-in zonotopes, mimicking the factorization structure of the correlators themselves.

We show that evaluating the canonical form of this polytope at the origin---equivalently, calculating the volume of the dual polytope---reproduces the flat-space correlator. This complements the findings of \cite{Figueiredo:2025daa}, which also introduced a polytope construction for the correlator.  Finally, in a simple example, we show that the wavefunction decomposition corresponds to a subdivision of the dual in-in zonotope.

\section{Correlator combinatorics}\label{sec:rev}
Our motivation comes from equal-time correlators for a polynomial scalar theory in flat space. These correlators are defined by integrating field insertions on a fixed time slice against the modulus square of the vacuum wavefunctional
\begin{align}
\langle \Phi_1 \ldots \Phi_n \rangle = \int \mathcal{D} \phi \ \Phi(0, \vec{x}_1) \ldots \Phi(0, \vec{x}_n) |\Psi[\phi]|^2.
\end{align}
The vacuum wavefunctional $\Psi[\phi]$ is defined by a path integral over field configurations that evolve from the vacuum in the distant past and end on a specified field configuration at the final time slice.

As with many observables in quantum field theory, these correlators are computed perturbatively as a sum over Feynman diagrams. We denote the contribution of a fixed diagram $G$ to the correlator by $\langle G \rangle$, and we will refer to these individual contributions simply as correlators.
\begin{figure}
\centering
\includegraphics[scale=0.9]{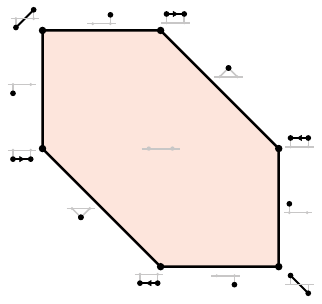}
\caption{The in-in zonotope $\mathcal{I}(P_2)$ with boundaries labelled by in-in diagrams. }
 \label{fig:cube_P2}
\end{figure}
\newline\newline
{\bf In-in formalism.}
In the in-in formalism \cite{Schwinger:1960qe,Keldysh:1964ud,feyn}, the correlator $\langle G \rangle$ is computed by assigning each vertex of the graph to either the forward or backward branches of the in-in contour. This is represented graphically by drawing the vertices of the graph either above or below the boundary, which is usually depicted as a horizontal line. The correlator then decomposes as a sum over all $2^{|V(G)|}$ forward/backwards vertex assignments. As a simple example, the two-vertex path graph, denoted $P_2$, admits the expansion
\begin{align}
\langle \raisebox{+0.02cm}{\includegraphics[scale=0.8]{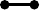}}  \rangle  &= \raisebox{-0.21cm}{\includegraphics[scale=0.9]{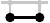}}+\raisebox{-0.21cm}{\includegraphics[scale=0.9]{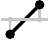}}+\raisebox{0.047cm}{\includegraphics[scale=0.9]{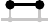}}+\raisebox{-0.21cm}{\includegraphics[scale=0.9]{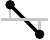}}.
\label{eq:two_pt}
\end{align}
{\bf Feynman rules.}
Each term appearing in \eqref{eq:two_pt} corresponds to a nested time-integral defined by a set of Feynman rules, see for instance \cite{Donath:2024utn}. However, since our aim is to let the diagrams themselves guide us toward an underlying geometry---rather than relying on details of the propagators---we set aside the specifics of the Feynman rules and focus instead on their combinatorial structure.

The grey edges are referred to as bulk-to-boundary propagators, these will serve only to distinguish the vertices of the graph $G$. Edges with end points on either side of the boundary correspond to Wightman propagators. These do not depend on the relative time ordering of the endpoints and hence remain unoriented. Edges between vertices on the same branch correspond to time or anti-time ordered propagators. These two contributions are represented graphically by orienting the corresponding edge in the graph, as demonstrated by
\begin{align}
 \raisebox{-0.1cm}{\includegraphics[scale=0.9]{figs/poly_prop}}= \raisebox{-0.21cm}{\includegraphics[scale=0.9]{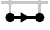}}+\raisebox{-0.21cm}{\includegraphics[scale=0.9]{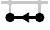}}.
\end{align}
Using these rules, the expansion of the correlator for the path graph on two vertices becomes 
\begin{align}
\langle \raisebox{+0.02cm}{\includegraphics[scale=0.8]{figs/pt2_bb}}  \rangle  =& \raisebox{-0.275cm}{\includegraphics[scale=0.8]{figs/poly_vert_1}}+\raisebox{-0.275cm}{\includegraphics[scale=0.8]{figs/poly_vert_2}}+\raisebox{-0.18cm}{\includegraphics[scale=0.8]{figs/poly_vert_3}}+\raisebox{0.067cm}{\includegraphics[scale=0.8]{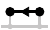}}+\raisebox{0.067cm}{\includegraphics[scale=0.8]{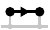}}+\raisebox{-0.18cm}{\includegraphics[scale=0.8]{figs/poly_vert_6}}.
\label{eq:two_pt_or}
\end{align}
For graphs containing cycles, only acyclic orientations of the graph survive in the correlator, for instance, the complete graph on three vertices has the expansion 
\begin{align}
\langle K_3  \rangle  =&\raisebox{-0.275cm}{\includegraphics[scale=0.55]{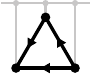}}+\raisebox{-0.22cm}{\includegraphics[scale=0.7]{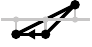}}+ \ldots,
\end{align}
where the ellipses indicate the remaining $22$ diagrams obtained via permuting the vertex labels and reflecting through the boundary.
\begin{table}[]
\centering
\begin{tabular}{c|c c c c c c c}
\hline
  $G$\textbackslash dim    & 0 & 1 & 2 & 3 & 4 & 5  \\
\hline
$P_5$ & 162 & 432 & 422 & 180 & 30 & 1 \\
$S_ 5$   & 162 & 448 & 464 & 216 & 40  & 1   \\
$C_ 5$   & 240 & 660 & 660 & 280 & 42 & 1   \\
$K_5$ & 720 & 1800 & 1560 & 540 & 62 & 1 \\
\hline
\end{tabular}
\caption{The number of in-in diagrams for the path, star, cycle and complete graph by number of vertices.}
\label{tab:counts}
\end{table}
 \newline\newline
{\bf Doubling up.}
\begin{figure}
\centering
\includegraphics[scale=0.55]{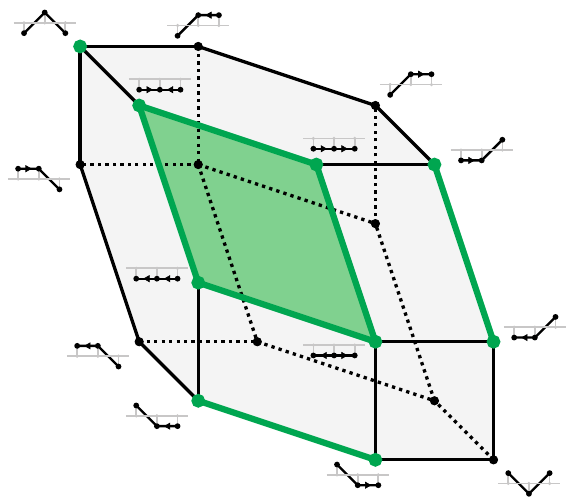}
\caption{The in-in zonotope $\mathcal{I}(P_2)$ with vertices labelled by in-in diagrams.}
 \label{fig:3d}
\end{figure} 
Upon evaluating the associated integrals, each diagram appearing in the in-in expansion contributes a rational function to the correlator. These rational functions are naturally expressed in terms of variables $(x_v,y_e)$ associated with the vertices and edges of the graph, and can be computed combinatorially, as discussed in \cite{Fevola:2024nzj,Glew:2025ugf,Glew:2025mry}. The graph variables are simply related to the kinematics of the physical process \cite{CosmoPolys}, although the precise details will not be important for our purposes. 

Notably, terms reflected through the boundary evaluate to the \emph{same} rational function, resulting in an overall factor of two in the correlator. For example, evaluating the integrals associated to the diagrams in \eqref{eq:two_pt_or} gives
\begin{align}
\langle P_2 \rangle  & \propto \frac{1}{(x_1+x_2)(x_1+y_{12})} + \frac{1}{(x_1+x_2)(x_2+y_{12})}  \notag \\
&+ \frac{1}{(x_1+y_{12})(x_2+y_{12})} + \frac{1}{(x_1+x_2)(x_1+y_{12})}  \notag \\
&+ \frac{1}{(x_1+x_2)(x_2+y_{12})} + \frac{1}{(x_1+y_{12})(x_2+y_{12})}.
\end{align}
However, as we will see in the next section, from a geometric perspective, it is natural to consider these diagrams independently.
\newline\newline
{\bf Generating diagrams.}
Anticipating the need to label higher dimensional faces of the in-in zonotope we introduce a procedure for generating new diagrams from old, each reducing the number of vertices by one. First, we may contract any oriented edge that results in a graph free of directed cycles. Note, any parallel edges in the resulting contracted graph are identified. Examples of this move are given by 
\begin{align}
\raisebox{-0.35cm}{\includegraphics[scale=0.6]{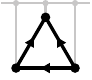}}  \rightarrow  \raisebox{-0.28cm}{\includegraphics[scale=0.6]{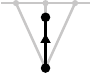}} \rightarrow \raisebox{-0.1cm}{\includegraphics[scale=0.8]{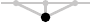}}.
\end{align}
Second, we may contract the grey edge connected to any {\it sink} vertex located below the boundary, or to any {\it source} vertex located above the boundary.  Here we define a sink/source vertex as one with no outgoing/incoming edges respectively. However, to avoid unnecessarily complicating the diagrams, we represent this operation by simply deleting the corresponding vertex. Examples of this move are the following 
\begin{align}
\raisebox{-0.15cm}{\includegraphics[scale=0.9]{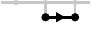}} \leftarrow \raisebox{-0.15cm}{\includegraphics[scale=0.9]{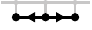}}  \rightarrow \raisebox{-0.15cm}{\includegraphics[scale=0.9]{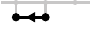}}.
\end{align}
We refer to the diagrams generated by these two moves collectively as {\it in-in diagrams}.
\newline\newline
{\bf Counting diagrams.}
For a fixed graph, it is instructive to count the number of in-in diagrams by number of vertices. Examples of these counts for the path, star, cycle and complete graph on five vertices are displayed in table~\ref{tab:counts}. Remarkably, their alternating sum vanishes, suggesting that they do indeed label the vertices of a polytope!

\section{in-in zonotopes}
In this section, we introduce a polytope, the {\it in-in zonotope}, whose faces are labelled by the in-in diagrams introduced above. \newline\newline
{\bf Facet description.}
Let $\mathfrak{g}$ be a connected, vertex-induced subgraph of $G$, denoted $\mathfrak{g} \preceq G$. Furthermore, let $x_v$ and $y_e$ be positive parameters associated to the vertices and edges of the graph. To each such subgraph we associate the following functions
\begin{align}
\alpha_\mathfrak{g} = \sum_{v \in V(\mathfrak{g})} \alpha_v, \quad  \quad  L_\mathfrak{g} = \sum_{v \in V(\mathfrak{g})} x_v + \sum_{e \in H(\mathfrak{g})} y_e,
\end{align}
where $H(\mathfrak{g})$ is the set of edges in $E(G)$ with exactly one endpoint in $\mathfrak{g}$. Then, in the vertex coordinates $\vec{\alpha}$, the in-in zonotope is defined as 
\begin{align}
\mathcal{I}(G) = \{ \vec{\alpha} &\in \mathbb{R}^{|G|} : |\alpha_{\mathfrak{g}}| \leq L_\mathfrak{g} \text{ for all } \mathfrak{g} \preceq G\},
\end{align}
where $|G| \equiv |V(G)|$ and the facets come in pairs due to the absolute value sign $|\alpha_{\mathfrak{g}}|$.
\newline\newline
{\bf Minkowski sum.}
Upon setting all $x_v=y_e=1$, the in-in zonotope is defined, for any graph $G$, as the following Minkowski sum 
\begin{align}
\mathcal{I}(G) = \mathcal{H}(G) \oplus \mathcal{Z}(G),
\end{align}
where the hypercube $\mathcal{H}(G)$ and graphical zonotope $\mathcal{Z}(G)$ are defined as the following sums of line segments 
\begin{align}
\mathcal{H}(G) = \bigoplus_{v \in V(G)} [-e_v,e_v], \quad \mathcal{Z}(G) = \bigoplus_{vv' \in E(G)} [-e_{vv'},e_{vv'}], \notag
\end{align}
with $e_{v}$ the unit vector in the $v^{\text{th}}$ coordinate direction and $e_{vv'} = e_v - e_{v'}$.
\begin{figure}
\centering
\includegraphics[scale=0.45]{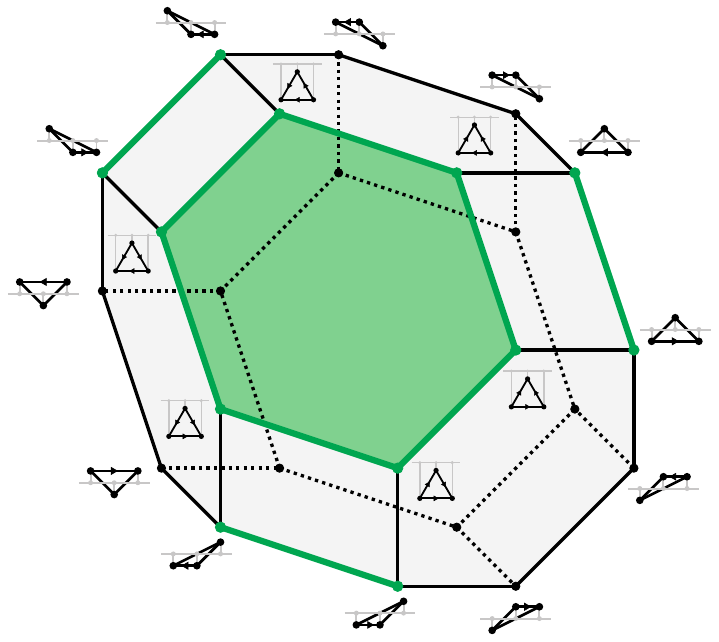}
\caption{The in-in zonotope $\mathcal{I}(K_3)$ with vertices labelled by in-in diagrams.}
 \label{fig:degen_poly_perm}
\end{figure} 
\newline\newline
{\bf Factorization.}
Consider the facet of $\mathcal{I}(G)$ defined by the hyperplane $F_{\mathfrak{g}} = \alpha_{\mathfrak{g}} + L_{\mathfrak{g}}=0$. This facet can be written as
\begin{align}
 \mathcal{I}(G)|_{F_{\mathfrak{g}}=0}= \mathcal{T}(\mathfrak{g})  \oplus \mathcal{Z}(\mathfrak{g})  \oplus \mathcal{I}(\bar{ \mathfrak{g}}),
\end{align}
where $\bar{\mathfrak g} = G \setminus \mathfrak g$ is the graph with $\mathfrak g$ removed, and the translation is
\begin{align}
\mathcal{T}(\mathfrak{g}) = -\sum_{v \in V(\mathfrak{g})}  x_v e_{v}-\sum_{vv' \in H(\mathfrak{g})}  y_{vv'} e_{vv'},
\end{align}
with the second sum taken over edges \(vv' \in H(\mathfrak{g})\) such that \(v \in V(\mathfrak g)\) and $v' \in V(\bar{\mathfrak{g}})$.

The factors $\mathcal{Z}(\mathfrak g)$ and $\mathcal{I}(\bar{\mathfrak g})$ lie in orthogonal coordinate subspaces, so their Minkowski sum takes on a product structure. For the same reason, $\mathcal{I}(\bar{\mathfrak g})$ factorizes over its connected components, giving the facet factorization
\begin{align}
 \mathcal{I}(G)|_{F_{\mathfrak{g}}=0} \simeq Z(\mathfrak{g})  \otimes \bigotimes_{\mathfrak{g}' \in \kappa(\bar{\mathfrak{g}})} \mathcal{I}(\mathfrak{g}'),
\end{align}
where $\kappa(\bar{\mathfrak{g}})$ denotes the connected components of $\bar{\mathfrak{g}}$.

As an example, consider the left-front hexagon of figure~\ref{fig:degen_poly_perm}, which satisfies $F_{2}=\alpha_{2}+x_{2}+y_{12}+y_{23}=0$. In this case, the boundary takes the following form
\begin{align}
 \mathcal{I}(K_3)|_{F_{2}=0} \simeq \mathcal{Z}(K_1) \times \mathcal{I}(K_2) \simeq \mathcal{I}(K_2),
\end{align}
where $\mathcal{Z}(K_1)$ is the graphical zonotope for a single vertex, a point, and $\mathcal{I}(K_2)$ is the in-in zonotope for a single edge, a hexagon in this case.
\newline\newline
{\bf Examples.}
The in-in zonotope for the path graph on two and three vertices are displayed in figure~\ref{fig:cube_P2} and~\ref{fig:3d} respectively. For the complete graph, the in-in zonotope is the permutohedron, whose three-dimensional example is displayed in figure~\ref{fig:degen_poly_perm}. More generally, $\mathcal{I}(G)$ is combinatorially equivalent to the graphical zonotope of the join graph $G * K_1$, formed by connecting a new vertex to every vertex in $G$.
\section{Correlators from Canonical forms}
The significance of the in-in zonotope extends beyond the combinatorial statements discussed so far. As we will demonstrate through examples, their canonical forms---once stripped of the measure and evaluated at the origin---reproduce the known results for scalar correlators.
\newline\newline
{\bf Examples.} Consider the polytope $\mathcal{I}(K_1) = [-x_1,x_1]$, with canonical form given by
\begin{align}
\omega = f_\omega \ d \alpha_1, \quad \quad f_\omega=\frac{1}{\alpha_1+x_1}- \frac{1}{\alpha_1-x_1}.
\label{eq:can_line}
\end{align}
Evaluating the canonical function $f_\omega$ at the origin we recover the known result for the correlator up to an overall factor
\begin{align}
\langle K_1 \rangle \propto f_{\omega}|_{{\alpha_1} =0} = \frac{2}{x_1}.
\label{eq:exam_K1}
\end{align}
Similarly, for the path graph on two vertices, the canonical function is given by
\begin{align}
f_{\omega} &= \frac{1}{(\alpha_1+\alpha_2+x_1+x_2)(\alpha_1+x_1+y_{12})} \notag \\
&+\frac{1}{(\alpha_1+\alpha_2+x_1+x_2)(\alpha_2+x_2+y_{12})} \notag \\
&+\frac{1}{(\alpha_1+x_1+y_{12})(\alpha_2+x_2+y_{12})}  + \ldots,
\end{align}
with the remaining terms given by replacing $\vec{\alpha} \rightarrow -\vec{\alpha}$ in the three terms above. Evaluating this function at the origin we find
\begin{align}
\langle P_2 \rangle \propto f_{\omega}|_{\vec{\alpha} = 0} = \frac{4(x_1+x_2+y_{12})}{(x_1+x_2)(x_1+y_{12})(x_2+y_{12})},
\label{eq:exam_P2}
\end{align}
which again produces the known result for the correlator up to an overall normalisation.
\section{Dual volumes}
\begin{figure}
\centering
\includegraphics[scale=0.9]{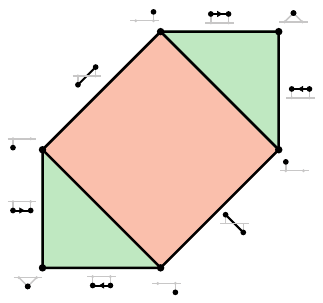}
\caption{The dual in-in zonotope $\mathcal{I}^\circ(P_2)$ subdivided according to the wavefunction decomposition.}
 \label{fig:dual}
\end{figure}

We now illustrate how the construction from the previous section, stripping the measure and evaluating the canonical function at the origin, is equivalent to computing the volume of the dual in-in zonotope \cite{Arkani-Hamed:2017tmz,gao2024dual}. We have checked this explicitly for all examples presented in the paper.
\newline\newline
{\bf Dual polytope.}
Since $\mathcal{I}(G)$ contains the origin, and we have its facet description, the dual polytope is obtained simply as the following convex hull
\begin{align}
\mathcal{I}^\circ(G) = \text{Conv}(\pm L^\circ_{\mathfrak{g}} | \ \mathfrak{g} \preceq G),
\end{align}
where we have defined the dual points
\begin{align}
L^\circ_{\mathfrak{g}} = \frac{1}{L_{\mathfrak{g}}} \sum_{v \in V(\mathfrak{g})} e_v.
\end{align}
As we now demonstrate, the (normalized) volume of this polytope directly computes the correlator.
\newline\newline
{\bf Examples.} 
Consider again the path graph on two vertices, the dual polytope in this case is depicted in figure~\ref{fig:dual}. The six points, starting from the bottom left and working clockwise are given by
\begin{align}
&p_1=-L^\circ_{12} && p_2=-L^\circ_{1}, &&&p_3=L^\circ_{2}, \notag\\
& p_4=L^\circ_{12}, && p_5=L^\circ_1, &&&p_6=-L^\circ_2,
\end{align}
where we have explicitly 
\begin{align}
 & L^\circ_{1} = \frac{(1,0)}{x_1+y_{12}}, &&L^\circ_{2} = \frac{(0,1)}{x_2+y_{12}}, &&&L^\circ_{12} = \frac{(1,1)}{x_1+x_2}. \notag
\end{align}
The area of this hexagon is easily computed by decomposing as
\begin{align}
\text{Vol}[612]+\text{Vol}[345]&= \frac{4y_{12}}{(x_1+x_2)(x_1+y_{12})(x_2+y_{12})}, \notag \\
\text{Vol}[2356]&= \frac{4}{(x_1+y_{12})(x_2+y_{12})}.
\label{eq:sub_hex}
\end{align}
One can check that summing these terms we recover the expression \eqref{eq:exam_P2}. 

As a final example, consider the graph $B_1$ consisting of a single vertex and loop edge. The correlator in this case is identical to \eqref{eq:can_line}. However, the additional edge defines natural internal points which may be used to subdivide the line as
\begin{align}
\mathcal{I}^\circ(B_1) = \includegraphics[scale=1]{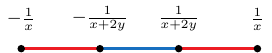}.
\end{align}
Calculating the length of the two colored regions we find
\begin{align}
\text{Vol}[\mathcal{I}^\circ(B_1)] = \frac{4y}{x(x+2y)}+\frac{2}{(x+2y)} = \frac{2}{x}.
\label{eq:sub_line}
\end{align}
Remarkably, the subdivisions of the in-in zonotope described in \eqref{eq:sub_hex} and \eqref{eq:sub_line} both match the wavefunction computation of their respective correlators!
\section{Conclusion}
Guided by the in-in formalism, we have  defined a family of polytopes---the {\it in-in zonotopes}---whose canonical forms encode individual graph contributions to flat-space scalar correlators. Our work suggests a number of interesting future research directions.   

We have seen in simple examples how the wavefunction decomposition of the correlator corresponds to subdivisions of the dual in-in zonotope. It would be interesting to study this systematically, and make connection to combinatorial formulas presented in \cite{Fevola:2024nzj,Glew:2025ugf,Glew:2025mry}. This  could provide a link to the polytope representation of the correlator discussed in \cite{Figueiredo:2025daa}.

A representation of the correlator that repackages the $2^{|G|}$ terms of the in-in expansion into a single integral was recently explored in \cite{Donath:2024utn,Arkani-Hamed:2025mce}; see also \cite{Chowdhury:2023arc,Sachs3} for related discussions. Likely, this integral admits an interpretation as an integral over the normal fan of the in-in zonotope, thereby providing a global Schwinger parametrization for correlators, analogous to constructions developed for scattering amplitudes \cite{Cachazo:2022vuo,Arkani-Hamed:2023lbd}. Clarifying this connection would be a natural next step.

The flat-space correlators discussed here serve as building blocks for cosmological correlators of conformally coupled scalars. These cosmological correlators satisfy a set of differential equations---the \emph{kinematic flow}---which admit a combinatorial formulation \cite{DiffEq_CosmoCorr,KinFlow}. Recently, it has been observed that the combinatorics of the kinematic flow for single graphs are captured by a collection of graphical zonotopes \cite{Baumann:2025qjx,Glew:2025ypb}. Since these graphical zonotopes also appear as faces of the in-in zonotope, it is natural to ask whether the in-in zonotopes also play a role in this context.

The in-in zonotopes capture contributions to the correlator coming from single Feynman diagrams. Recently, it was shown that the sum over all tree-level graphs for $\text{tr}(\phi^3)$ theory can be simultaneously encoded by a single polytope, the cosmohedron \cite{Cosmohedron,Glew:2025otn}. This raises the question of whether an analogous construction exists for our polytopes.

\acknowledgments{We would like to thank Stefan Forcey, Francisco Vazao, Nima Arkani-Hamed and Carolina Figueiredo for related discussions.}

\onecolumngrid

\bibliographystyle{apsrev4-1.bst}
\bibliography{Refs.bib}

\end{document}